\documentclass[final,5p,times,twocolumn,authoryear]{elsarticle}

\usepackage{amssymb}
\usepackage{lipsum}
\usepackage{amsmath}
\usepackage[table]{xcolor}
\usepackage{subcaption}
\usepackage[hyphens]{url}

\journal{Journal of a Cleaner Production }

\begin{document}

\begin{frontmatter}




\title{Beyond 2050: From deployment to renewal of the global solar and wind energy system} 


\author[inst1]{Joseph Le Bihan\corref{correspondingauthor}}
\ead{joseph.le-bihan@etu.u-paris.fr}
\cortext[correspondingauthor]{Corresponding author. : J. Le Bihan, Université Paric Cité, Bâtiment Condorcet, 10 rue Alice Domon et Léonie Duquet, 75013, Paris, France}

\affiliation[inst1]{organization={Université Paris Cité, CNRS, LIED UMR 8236},
            city={Paris},
            postcode={F-75006},
            country={France}}

\author[inst1]{Thomas Lapi}
\ead{thomas.lapi@etu.u-paris.fr}
\author[inst1]{José Halloy}
\ead{jose.halloy@u-paris.fr}

\begin{abstract}

The global energy transition depends on large-scale photovoltaic (PV) and wind power deployment. While 2050 targets suggest a transition endpoint, maintaining these systems beyond mid-century requires continuous renewal, marking a fundamental yet often overlooked shift in industrial dynamics.
This study examines the transition from initial deployment to long-term renewal, using a two-phase growth model: an exponential expansion followed by capacity stabilization. By integrating this pattern with a Weibull distribution of PV panel and wind turbine lifespans, we estimate the annual production required for both expansion and maintenance.
Our findings highlight two key factors influencing production dynamics: deployment speed and lifespan. When deployment occurs faster than the average lifespan, production overshoots and exhibits damped oscillations due to successive installation and replacement cycles. In contrast, gradual deployment leads to a smooth increase before stabilizing at the renewal rate. Given current scenarios, the PV industry is likely to experience significant oscillations—ranging from 15\% to 60\% of global production—while wind power follows a monotonic growth trajectory. These oscillations, driven by ambitious energy targets, may result in cycles of overproduction and underproduction, affecting industrial stability.
Beyond solar and wind, this study underscores a broader challenge in the energy transition: shifting from infrastructure expansion to long-term maintenance. Addressing this phase is crucial for ensuring the resilience and sustainability of renewable energy systems beyond 2050.

\end{abstract}




\end{frontmatter}




\section{Introduction}
\label{introduction}

Energy transition goals call for a significant deployment of renewable energy (RE) options. Solar photovoltaic (PV) and wind power technologies are currently one of the most promising primary energy options to achieve decarbonization goals (in addition to biofuels, hydroelectricity, and geothermal energy). Both the installed capacity of photovoltaic and wind power technology has increased significantly in the last twenty years [\cite{Histdata}] in response to climate and renewable energy policies around the world [\cite{Ieapolicydata}]. Numerous scenarios investigating the deployment of solar and wind energy, along with their potential, have been conducted across various geographical scales and under different constraints [\cite{cherp2021,joshi2021}]. In addition, the deployment of PV and Wind technologies has been studied from other perspectives, including cost reduction challenges [\cite{cost1,cost2}], material supply issues [\cite{mat1,mat2}], social acceptance [\cite{social1,social2}], and more [\cite{Choudhary2019,allouhi2022}].

A common feature shared by most renewable energy forecast studies is the 2050 target, as a result of the carbon neutrality goals set by numerous countries for 2050 [\cite{Ieadata,Ieapolicydata}]. While such approaches enable us to pinpoint significant challenges during the deployment phase of PV or wind technologies, the post-2050 phase remains a blind spot, especially in terms of installed capacity renewal. In other words, most analysis focus on the deployment phase [\cite{Ieadata,Bogdanov}], while the long-term renewal of the solar and wind energy system beyond 2050 has been little studied. 

The question of renewal has major policy implications, because PV and Wind industries are entangled in a complex set of societal dynamics. Solar and wind energy systems includes issues as broad as : labour force, know-how, materials supply, commodity markets, regulations, etc. The question of these systems renewal therefore raises critical strategic challenges for policymakers and requires an understanding of RE technologies renewal dynamics and their retroactive effects on the entire RE system.

This study focuses on the renewal issues of PV and Wind power technology beyond 2050, based on the deployment rate and lifespan of both PV and Wind power. It discusses the results of the model and highlights strategic implications from an interdisciplinary perspective.

\subsection*{Objectives of this analysis}

This study aims to analyze the deployment of PV and wind systems required to meet the scenarios outlined by the International Energy Agency (IEA) [\cite{Ieadata}] from a long-term perspective. To address this, a concise mathematical model is developed to represent the temporal dynamics of photovoltaic and wind power deployment, leveraging historical data and projections based on IEA scenarios. The streamlined design of the model facilitates a clear understanding of the primary challenges, avoiding the opacity often associated with highly complex models. This makes it a valuable tool for effectively analyzing long-term dynamics.

Rather than providing precise forecasts, this study takes a forward-looking approach to assessing the scale of renewal needs and the dynamics of the transition from a deployment-driven industry to a renewal-driven industry. Thanks to its clear and simple approach, this study makes it easier to understand long-term developments. It also provides an opportunity to analyze the potential impact of industrial decisions, such as extending the lifespan of solar panels or accelerating deployment.

\section{Modeling installation and renewal of the world solar and wind energy capacity}
\label{modeling}

The question of item production required to support the deployment of a global energy technology fleet (solar or wind) is well addressed in the literature. This analysis typically involves defining a scenario for active solar/wind capacity (in TW) and calculating its rate of change over time, which corresponds to the annual panel/wind turbine production (in TW/year).
    
The idea of 'renewal production' adds an extra layer of complexity to this scenario. The initial deployment directly affects the renewal production requirements. If all panels/wind turbines are installed in one year, the renewal demands will be negligible for the next 30 years. Furthermore, the lifespan of panels/wind turbines does not have a fixed value, but rather follows a probability distribution. Factors other than physical wear and tear, such as economic considerations, can also influence the lifespan of technological equipment. These factors could vary significantly between solar panels and wind turbines.
    
\begin{figure*}[h]
    \centering
    \includegraphics[width=0.85\linewidth]{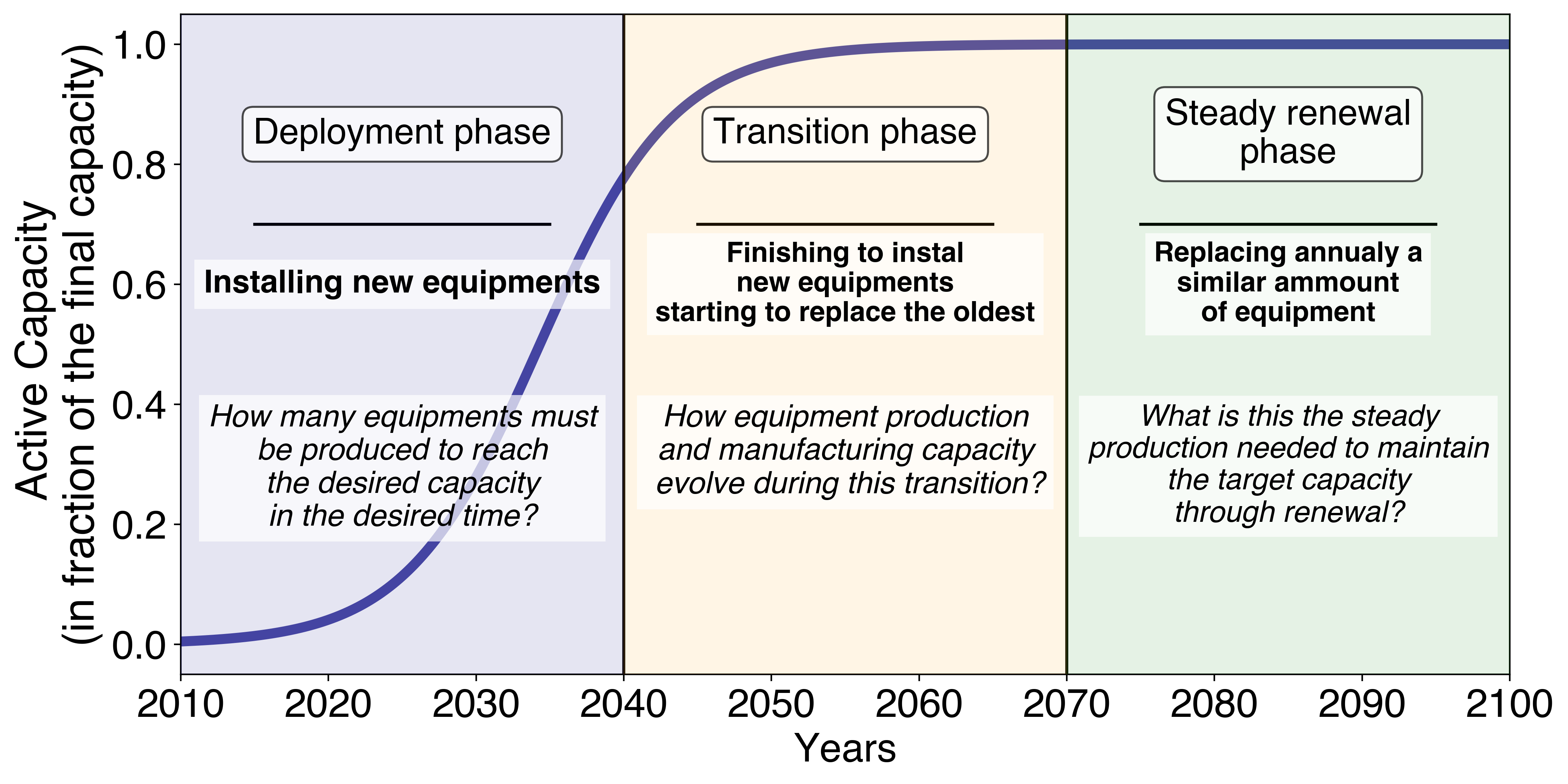}
    \caption{\textbf{Trajectory and challenges in the deployment and renewal of the global photovoltaic (PV) industry} This figure divides the global PV deployment dynamic, with the blue curve representing active PV global capacity, into three phases. The deployment phase (now - 2040) where the dynamic is driven by the production and installation of new panels. The transition phase (2040 - 2070) where the dynamic is driven both by the last installations and first renewals. The steady renewal phase (2070 - future) where the dynamic is driven by renewals of panels, and where the successive renewals have resulted in a stabilized panel production.}
    \label{fig:Intro}
    
\end{figure*}

Figure \ref{fig:Intro} provides a graphical illustration of the interplay between deployment and renewal in terms of capacity and production. This representation is applied to PV; however, it would be exactly the same for wind energy.

\subsection{Scenario assumptions}

Our model operates under specific boundary conditions that justify an explicit definition. We adopt a 'business-as-usual' framework where, all other things being equal, economic, industrial, energy and material constraints are deliberately excluded from the deployment and production of photovoltaic installations.

This foresight approach isolates and examines the pure effects of solar/wind park renewal dynamics. The resulting production scenarios should thus be interpreted as indicators of panel replacement patterns, independent of external factors.

Fundamentally, this represents a physical model investigating how two key constraints — the maintenance of active photovoltaic/wind capacity and panel/wind turbine service lifetimes — shape annual equipment production. Rather than providing precise predictions, our results highlight a limiting case that reveals risks and opportunities previously ignored in the scenarios of the current literature.

This targeted approach opens the door to broader investigations, particularly into how the identified effects may interact with the various constraints intentionally excluded from our analysis. By adopting the assumption 'all else is equal', we isolate these core dynamics while recognizing that real-world scenarios will inevitably involve complex interactions with additional factors.

\subsection{Minimal model for solar and wind capacity deployment}
\label{sec:MinimalModel}

This section introduces the parsimonious mathematical model employed to couple deployment and renewal dynamics in PV or Wind power production. A detailed presentation and validation of this model have been conducted in the context of two different industries: Smartphones and Nuclear [\cite{LeBihan2025}]. Here, we present the core aspects of the model adapted to the renewable energy technologies.

{\bf Parameters:}
\begin{itemize}
    \item $Capacity\,(t)$ - the active global PV capacity / wind power capacity over time, measured in TW. In this model, a \textit{logistic curve} (S-curve) is fitted to the milestones of the deployment scenario (see Figure \ref{fig:DepScenario}). The logistic curve is characterized by the following three parameters:
    \begin{itemize}
        \item $K$ - The target or final capacity in TW, representing the plateau value of the \textit{S-curve} after the deployment phase.
        
        \item $\tau_{\text{dep}}$ - The characteristic deployment time in years, defining the rate of deployment. It is specific to the logistic curve used (see Eq. \ref{eq:capa}).
        
       \item $t_{peak}$ - Usually, a logistic curve is defined by two parameters ($\tau_{\text{dep}}$, $K$) and an initial condition . Here, we fit the IAE scenarios with three parameters, avoiding the arbitrary choice of a point as the initial condition. The peak time, $t_{peak}$, depends on the temporal origin of the deployment. This parameter has no influence on overall system dynamics.
        
    \end{itemize}

    \item $p_{EoL}\, (\theta)$ — The probability density function of the panel/wind turbine lifespan, representing the likelihood that a panel/wind turbine reaches the end of life (EoL) $\theta$ years after production, is considered. The specific causes of EoL, whether technical (e.g., wear and degradation) or socio-economic (e.g., replacement by higher-efficiency panels), are discussed in this study. A Weibull distribution, commonly used in the literature, is used and parameterized accordingly:
    \begin{itemize}
        \item $\tau_{\text{EoL}}$ - The average lifespan of the equipment in years.
        \item $CV_{EoL} = \sigma_{EoL}/\tau_{\text{EoL}}$ - The coefficient of variation, representing the variance of lifespan normalized by the average lifespan.
    \end{itemize}
\end{itemize}

\begin{equation}
\label{eq:capa}
    Capacity\,(t) = \frac{K}{1+e^{-\left(t-t_{peak}\right)/\tau_{\text{dep}}}}    
\end{equation}

\begin{equation}
\label{eq:eol}
    p_{EoL}\, (\theta) = \frac{k}{\lambda} \left(\frac{\theta}{\lambda}\right)^{k-1} \exp\left(-\left(\frac{\theta}{\lambda}\right)^k\right)
\end{equation}

The Weibull distribution is parameterized by a scale parameter $\lambda$ and a shape parameter $k$, which relate to the aforementioned parameters as follows:
$$ \left(\tau_{\text{EoL}}, CV_{EoL}\right) = \left(\lambda \Gamma(1+\frac{1}{k}), \sqrt{\frac{\Gamma(1+\frac{2}{k})}{\Gamma(1+\frac{1}{k})^2} - 1}\right) $$

{\bf Model Equations:}

The model defines production as the sum of two dynamical processes:

\begin{itemize}
    \item $P_{dep}\,(t)$ - The \textit{deployment} production, measured in TW/year, representing panel/wind turbine production to increase active capacity to the planned level for the next year (see Eq. \ref{eq:dep}).
    \item $P_{renew}\,(t)$ - The \textit{renewal} production, measured in TW/year, representing panel/wind turbine production to replace those reaching EoL. This includes all panels/wind turbine produced in previous years that reach EoL at time $t$ (see Eq. \ref{eq:renew}).
\end{itemize}

The total production is expressed as the sum of these two components:
\begin{align}
    &P_{tot}(t) = P_{dep}(t) + P_{renew}(t)\label{eq:tot}\\
    &P_{dep}(t) = \frac{d}{dt}Capacity(t) = \frac{K}{\tau_{\text{dep}}} \cdot \frac{e^{-\left(t-t_{peak}\right)/\tau_{\text{dep}}}}{\left(1+e^{-\left(t-t_{peak}\right)/\tau_{\text{dep}}}\right)^2}\label{eq:dep}\\
    &P_{renew}(t) = \int_{0}^{\infty}P_{tot}(t-\theta)p_{EoL}(\theta)d\theta\label{eq:renew}
\end{align}

The equation for deriving annual production over time is therefore:
\begin{equation}
\label{eq:model}
    P_{tot}(t) = \frac{d}{dt}Capacity(t) + \int_{0}^{\infty}P_{tot}(t-\theta)p_{EoL}(\theta)d\theta
\end{equation}

\subsection*{Computational methods}

Numerical simulation, equation solving and visualization are made within a Python environment and the packages Numpy, Scipy and Matplotlib.

Equation \ref{eq:model} has been discretized in the following way :
\begin{multline}
    P_{tot}[t] = \left(Capacity[t] - Capacity[t-\delta_{t}]\right)/\delta_{t} \\
    + \Sigma_{\theta = 1}^{100}P_{tot}[t-\theta](F_{EoL}(\theta)-F_{EoL}(\theta-\delta_{t}))
\end{multline}

Where $F_{EoL}$ represents the cumulative distribution function of $p_{EoL}$ that can be computed analytically. A time step of $\delta_{t}=0.1 \text{years}$ has been used, and the probability density $p_{EoL}$ has been considered null after $100$ years.

\subsection{Scenarios for RE Deployment and Technologies Lifespan Distributions}
\label{modeldata}

This section outlines the two key parameters required for the model: active capacity deployment and technology lifespan distributions. The methods for selecting or calculating these parameters from the literature are detailed below.

\subsubsection{Historical Deployment and Deployment Scenarios}

\begin{figure}[!h]
    \centering
    \includegraphics[width=\linewidth]{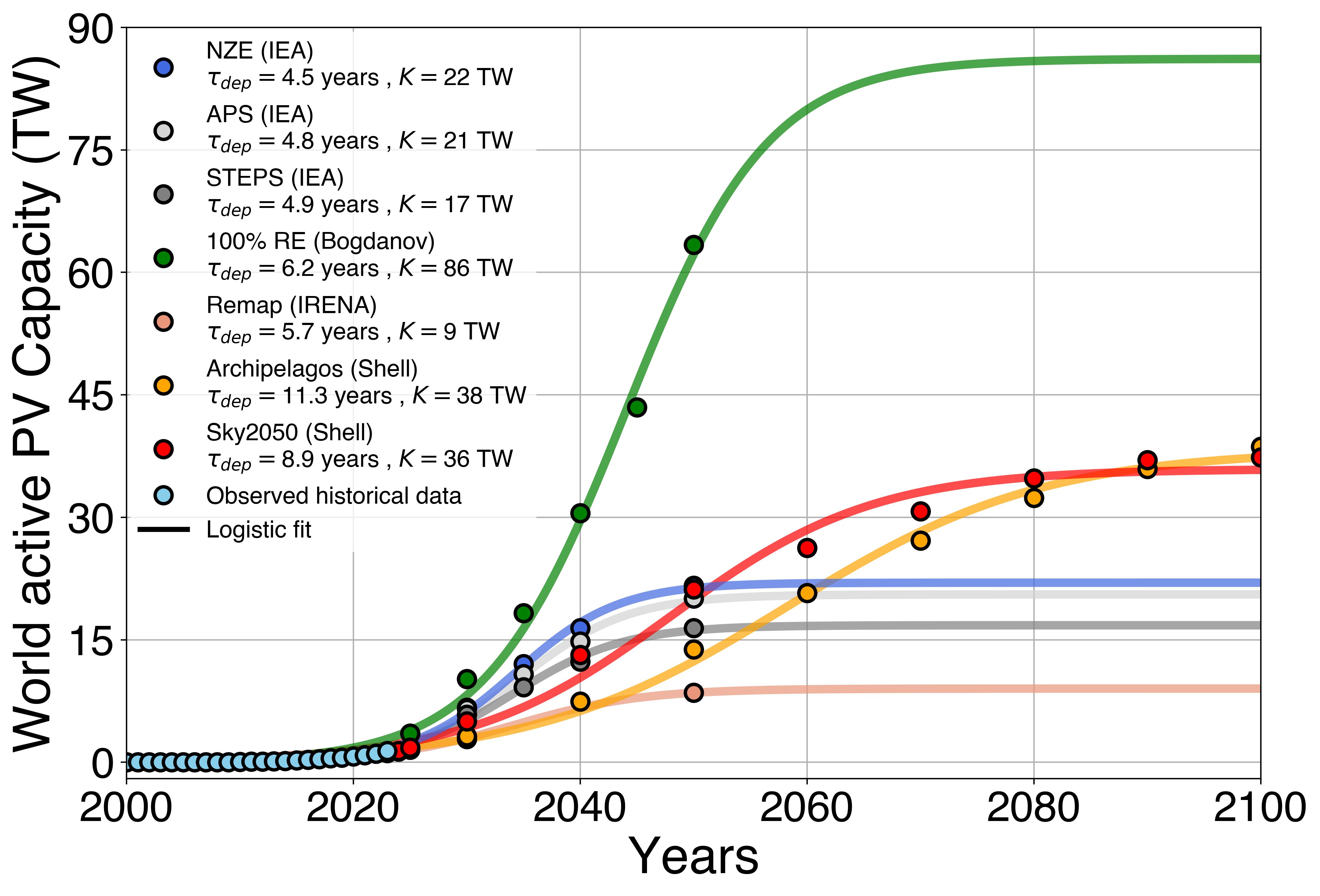}
    \caption{{\bf Projections for global solar PV deployment.} Historical data and future projections of global PV capacity (TW) from 2000 to 2100 are shown. These are modeled using various S-curves (Logistic) based on deployment scenarios described in Table \ref{tab:PVscenarios}.}
    \label{fig:DepScenarioPV}
\end{figure}

\begin{figure}[!h]
    \centering
    \includegraphics[width=\linewidth]{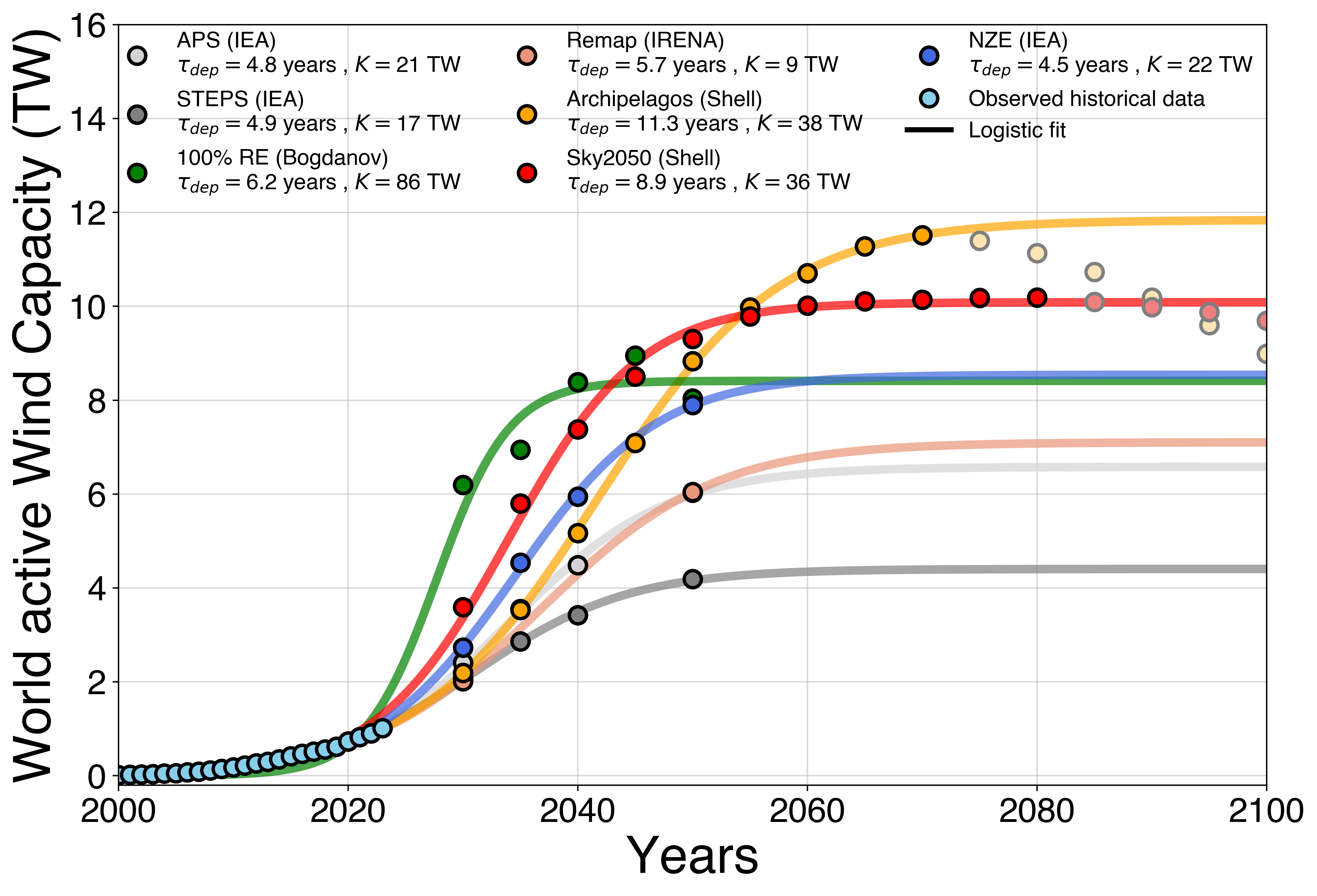}
    \caption{{\bf Projections for global wind power deployment.} Historical data and future projections of global Wind power capacity (TW) from 2000 to 2100 are shown. These are modeled using various S-curves (Logistic) based on deployment scenarios described in Table \ref{tab:PVscenarios}.}
    \label{fig:DepScenarioWind}
\end{figure}

As described in Section \ref{sec:MinimalModel}, the global active solar/wind capacity is modeled using a logistic curve fitted to historical data and future projections from selected deployment scenarios. Figures \ref{fig:DepScenarioPV} and \ref{fig:DepScenarioWind} illustrates the resulting fits for various scenarios detailed in Table \ref{tab:PVscenarios}.

\begin{figure}[!h]
    \centering
    \includegraphics[width=\linewidth]{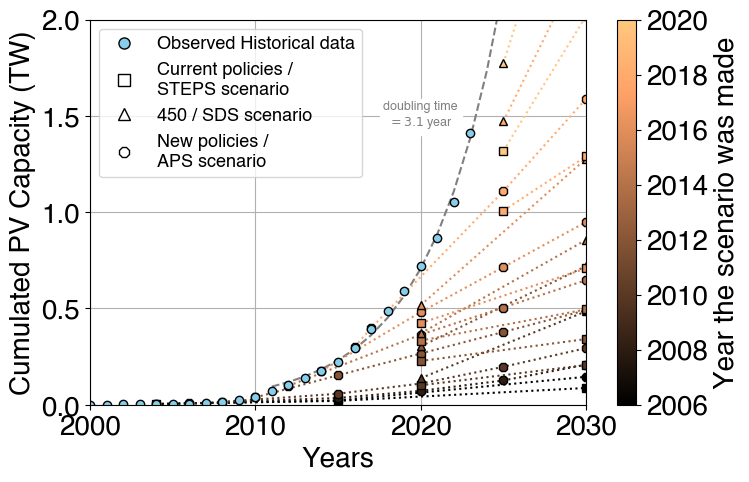}
    \caption{{\bf Historical deployment of solar PV capacity.} Historical data of global PV capacity (TW) from 2000 to 2023 is shown alongside older IEA projections (preceding those presented in Table \ref{tab:PVscenarios}).}
    \label{fig:PVexponential}
\end{figure}

The choice of a sigmoid function, such as the logistic curve, is supported by literature on technological deployment and corroborated by historical data. It is noteworthy that previous IEA scenarios consistently underestimated the exponential growth of solar PV capacity (see Figure \ref{fig:PVexponential}).

In addition to IEA scenarios, this study incorporates scenarios from the International Renewable Energy Agency (IRENA), Shell and an ambitious 100\% renewable energy scenario from the scientific literature [\cite{Bogdanov}] (see Table\ref{tab:PVscenarios}). These scenarios envision higher target capacities, with some projections extending to 2100 rather than 2050. While not exhaustive, these examples provide insights into the range of expected deployment trajectories.

The assumption of steady capacity from 2050 onwards may seem a little pessimistic. However, the date of 2050 is not particularly important, as this study shows that industrial dynamics depend essentially on deployment speed and the energy equipment lifespan. The date at which capacity reaches saturation, corresponding to the transition from a deployment regime to a renewal regime, has no impact on the dynamics. 

The central assumption of this model is the eventual occurrence of saturation at some point. While a detailed justification of this hypothesis lies beyond the scope of this article, several underlying factors can be outlined, including the stabilization of energy demand across various socio-economic trajectories, diminishing returns as optimal locations are utilized, and constraints associated with integrating photovoltaic systems into the power grid.

In this study, we focus on the NZE scenario from the IEA, which serves as a reference for both solar and wind energy. However, given the IEA’s repeated underestimation of PV deployment trends, this scenario may even be considered conservative for solar energy. Additionally, we analyze more ambitious scenarios: the 100RE scenario for solar energy and Shell’s Sky2050 scenario for wind energy. In all cases, the figures clearly illustrate the final capacity and the characteristic deployment time of each scenario, providing a transparent basis for assessing the associated dynamics.

\begin{table*}[h!]
\centering
\renewcommand{\arraystretch}{1.2}
\begin{tabular}{|| p{0.13\linewidth} || p{0.65\linewidth} | p{0.14\linewidth} ||} 
 \hline
 Name & Description & Source  \\ [0.5ex] 
 \hline\hline
 {\bf Historical Data} & Global solar (PV) and Wind (On-shore + Off-shore) capacity data compiled by IRENA & \cite{Histdata} \\ \hline
 {\bf NZE} & Net Zero Emissions : Achieves net-zero $\textrm{CO}_2$ emissions by 2050 and limits global temperature rise to +1.5°C & \cite{Ieadata} \\ \hline
 {\bf APS} & Announced Pledged Scenario: Based on announced national targets and ambitions & \cite{Ieadata} \\ \hline
 {\bf STEPS} & Stated Policies Scenario: Reflects the prevailing direction of energy system development & \cite{Ieadata}\\ \hline
 {\bf Archipelagos} & Focuses on energy security with innovation driven by competition, stabilizing global temperature rise at 2.2°C by 2100 & \cite{Shelldata} \\ \hline
 {\bf Sky2050} & Targets long-term climate security with net-zero emissions by 2050, stabilizing global temperature rise at +1.2°C & \cite{Shelldata}\\ \hline
 {\bf Remap} & Explores pathways to double the share of renewable energy in the global energy mix by 2030 & \cite{Irenadata} \\ \hline
 {\bf 100RE} & Achieves 100\% renewable energy system by 2050 & \cite{Bogdanov} \\ 
 \hline
\end{tabular}
\caption{{\bf Scenarios for global RE technologies deployment}. Overview of the different PV and Wind power deployment scenarios analyzed in this study (see Figure \ref{fig:DepScenarioPV} and \ref{fig:DepScenarioWind}). }
\label{tab:PVscenarios}
\end{table*}

\subsubsection{RE equipment End-of-Life Distribution}

\begin{figure}[h]
    \centering
    \includegraphics[width=\linewidth]{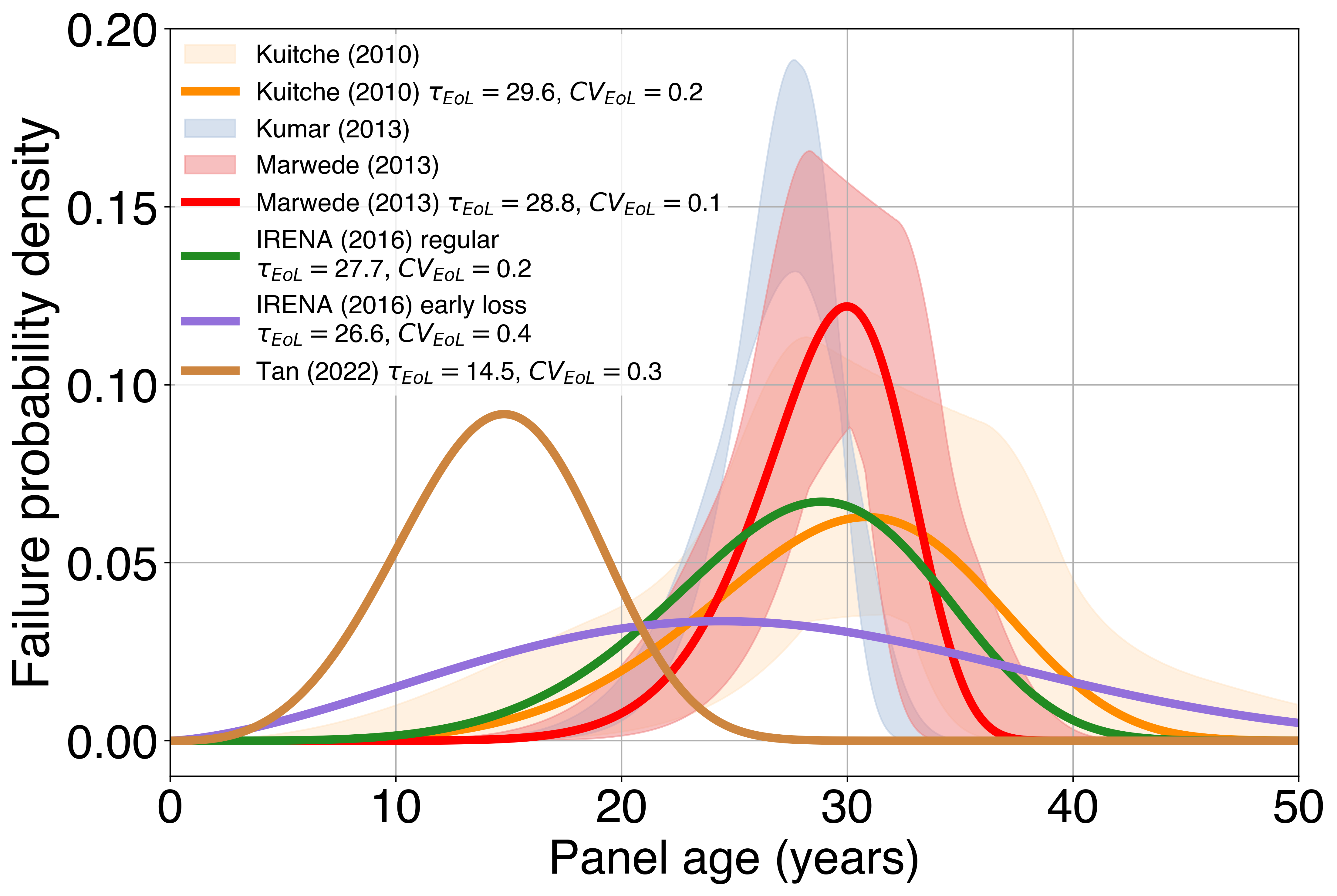}
    \caption{{\bf Distribution of solar panel end-of-life.} The figure displays various probability densities for panel lifetimes (see Table \ref{tab:PVEoLDistrib}). Most estimates suggest an average lifespan of 30 years based on degradation rates. The distribution from Tan reflects a shorter lifespan (15 years) influenced by economic factors.}
    \label{fig:PanelEol}
\end{figure}

\begin{figure}[h]
    \centering
    \includegraphics[width=\linewidth]{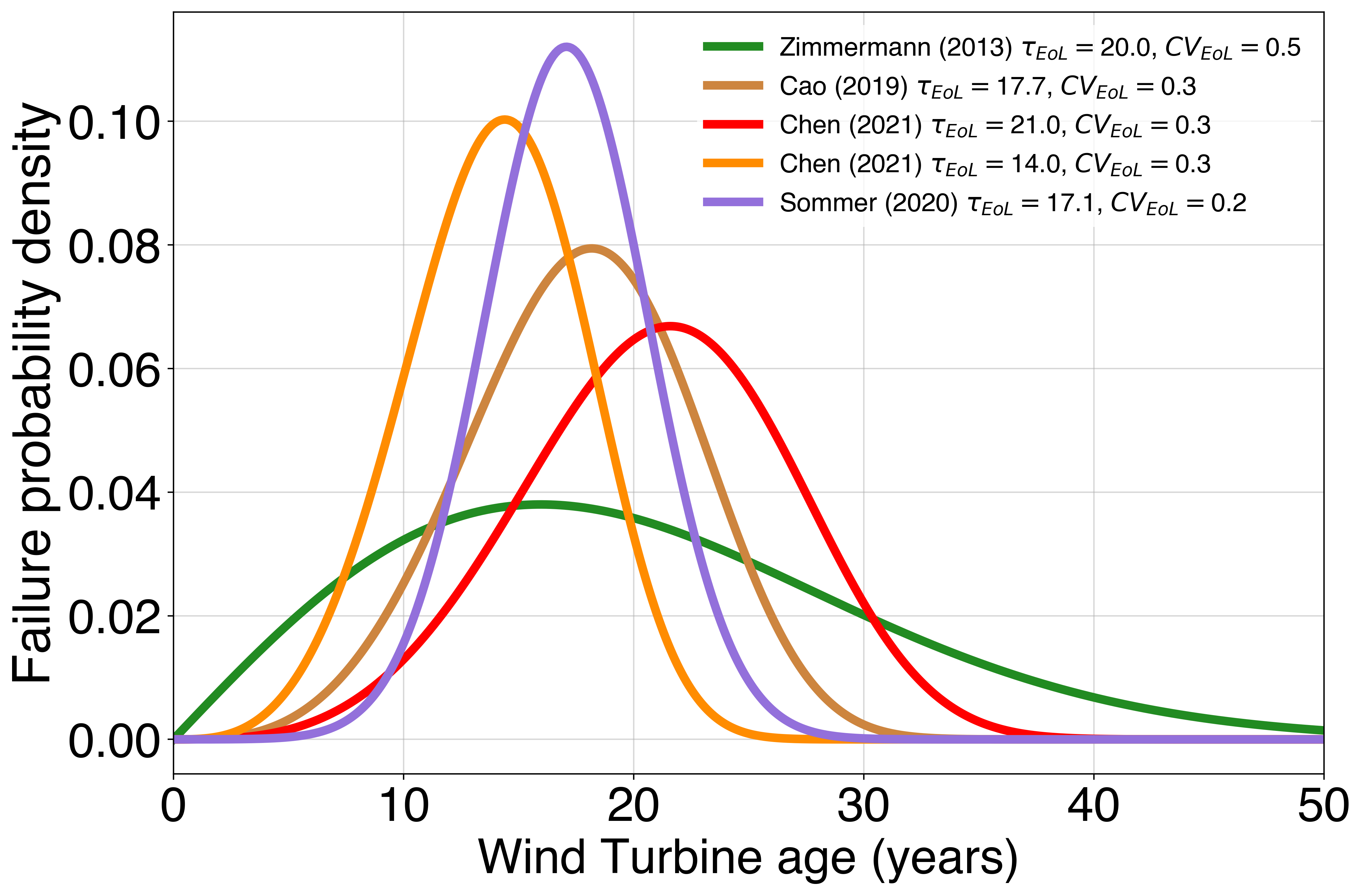}
    \caption{{\bf Distribution of wind turbine end-of-life.} The figure displays various probability densities for wind turbine lifetimes (see Table \ref{tab:WindEoLDistrib}). Most estimates suggest an average lifespan of 20 years.}
    \label{fig:WindTurbineEol}
\end{figure}

The \textit{renewal} dynamics are strongly influenced by the distribution of panel/wind turbine lifespans (Eq. \ref{eq:renew}), particularly by the ratio between the characteristic deployment time ($\tau_{\text{dep}}$) and the average item lifespan ($\tau_{\text{EoL}}$). Figure \ref{fig:PanelEol} and \ref{fig:WindTurbineEol} illustrates the EoL distributions reported in the literature, along with corresponding $\tau_{\text{EoL}}$ values.

Figure \ref{fig:PanelEol} shows a wide disparity in the panel lifespan distributions proposed in the literature. While most agree on an average life expectancy of around 30 years, the dispersion varies considerably. This 30-year value seems to correspond to a degradation effect on output power. Indeed, most panel manufacturers generally guarantee an output power of at least 85\% at 25 or 30 years. However "abrasions, scratches, stains, mechanical wear, rust, mildew, color difference and optical attenuation, normal wear and tear during use" are typically not covered [\cite{TongweiWarranty}] (see Table \ref{tab:PVEoLDistrib}).

The distribution provided in \cite{Tan2022} (brown curve in Figure~\ref{fig:PanelEol}) incorporates economic and failure-related factors, resulting in an average lifespan nearly half as long as the others. For the remainder of this study, the analysis will focus on this distribution and the one from \cite{Irena2016}, which is the most widely referenced in the literature. These two distributions will be used to evaluate the impact of effective lifespan before replacement on system dynamics, regardless of the underlying reason for replacement.

For wind turbines, the lifetime distributions appear to be more concentrated around a mean value of 18 years, with a less pronounced standard deviation variability compared to PV. This is primarily due to the use of a statistical approach based on actual decommissioning data from Denmark and Germany, in contrast to PV. In this study, we adopt the distribution used in \cite{cao2019}, which is derived from empirical data and appears to be widely accepted in the literature

\subsubsection*{RE equipment End-of-Life Distribution Data} 
\begin{table*}[h!]
\centering
\renewcommand{\arraystretch}{1.2}
\begin{tabular}{|| p{0.15\linewidth} || p{0.6\linewidth} | p{0.15\linewidth} ||} 
 \hline
 Name & Description & Source  \\ [0.5ex] 
 \hline\hline
 {\bf Kuitche (2010)} & Extrapolation of power output for panels installed over 8 years &\cite{Kuitche2010} \\ \hline
 {\bf Kumar (2013)} & Survival cycle testing of panel batches & \cite{Kumar2013} \\ \hline
 {\bf Marwede (2013)} & Combination of "early returns" due to defects/damages during transport and "end-of-life" due to power degradation &\cite{Marwede2013} \\ \hline
 {\bf IRENA (2016)} & Includes "regular" EoL based on Kuitche (2010) and "early loss" due to transport/installation issues & \cite{Irena2016} \\ \hline
 {\bf Tan (2022)} & Combines probabilities for "power decrease", "technical failure," and "economic motivations" & \cite{Tan2022}\\
 \hline
\end{tabular}
\caption{{\bf Panel lifespan distributions estimates}. Overview of the different PV lifespan distribution analyzed in this study (see Figure \ref{fig:PanelEol}). }
\label{tab:PVEoLDistrib}
\end{table*}

\begin{table*}[h!]
\centering
\renewcommand{\arraystretch}{1.2}
\begin{tabular}{|| p{0.17\linewidth} || p{0.54\linewidth} | p{0.17\linewidth} ||} 
 \hline
 Name & Description & Source  \\ [0.5ex] 
 \hline\hline
{\bf Zimmermann (2013)} & Assumptions and analogy with long living infrastructures & \cite{zimmermann2013} \\ \hline
 {\bf Cao (2019)} & Statistical analysis on Danish empirical data & \cite{cao2019} \\ \hline
 {\bf Chen (2021)} & Statistical analysis on Danish empirical data combine with a high and low replacement rate scenario & \cite{chen2021} \\ \hline
 {\bf Sommer (2020)} & Statistical analysis on German empirical data  & \cite{sommer2020}\\
 \hline
\end{tabular}
\caption{{\bf Wind Turbine lifespan distributions estimates}. Overview of the different Wind Turbine lifespan distribution analyzed in this study (see Figure \ref{fig:WindTurbineEol}). }
\label{tab:WindEoLDistrib}
\end{table*}

\newpage

\section{Results}

\subsection{The two potential deployment-renewal dynamics}
\label{sec:twodynamics}
The annual production patterns predicted by the model exhibit two distinct regimes - oscillating and monotonic - determined by the deployment scenario and the lifespan distribution of the RE equipment. The key parameter governing these profiles is the ratio $\tau_{\text{dep}}/\tau_{\text{EoL}}$, which represents the deployment rate (or time required to reach the target PV/Wind power capacity) relative to the average panel/wind turbine lifespan [\cite{LeBihan2025}]. A \textbf{critical threshold} for this ratio is approximately 0.3, corresponding to the installation of 70\% of the target capacity within the average lifespan of the equipment.

\begin{itemize}
    \item When the ratio $\tau_{\text{dep}}/\tau_{\text{EoL}}$ remains below the critical threshold, the production system transitions into a regime marked by significant overshoot followed by damped oscillations. We refer to this regime as \textbf{fast deployment}, arising from the compressed timeframe required to achieve the target capacity, which requires substantial overproduction relative to the renewal rate. As a result, successive replacement peaks occur, driven by the clustering of initial installations within a timeframe that is short compared to the average lifespan of the panels or wind turbines.

    \item Conversely, when the ratio $\tau_{\text{dep}}/\tau_{\text{EoL}}$ exceeds the critical threshold, the system enters a regime referred to as \textbf{slow deployment}. In this regime, production gradually scales to align with the renewal rate, avoiding overproduction. The extended duration of the initial deployment results in a more even distribution of installations, eliminating pronounced replacement peaks and naturally smoothing renewal demands over time.
\end{itemize}

\begin{figure*}[!h]
    \centering
    \includegraphics[width=0.9\linewidth]{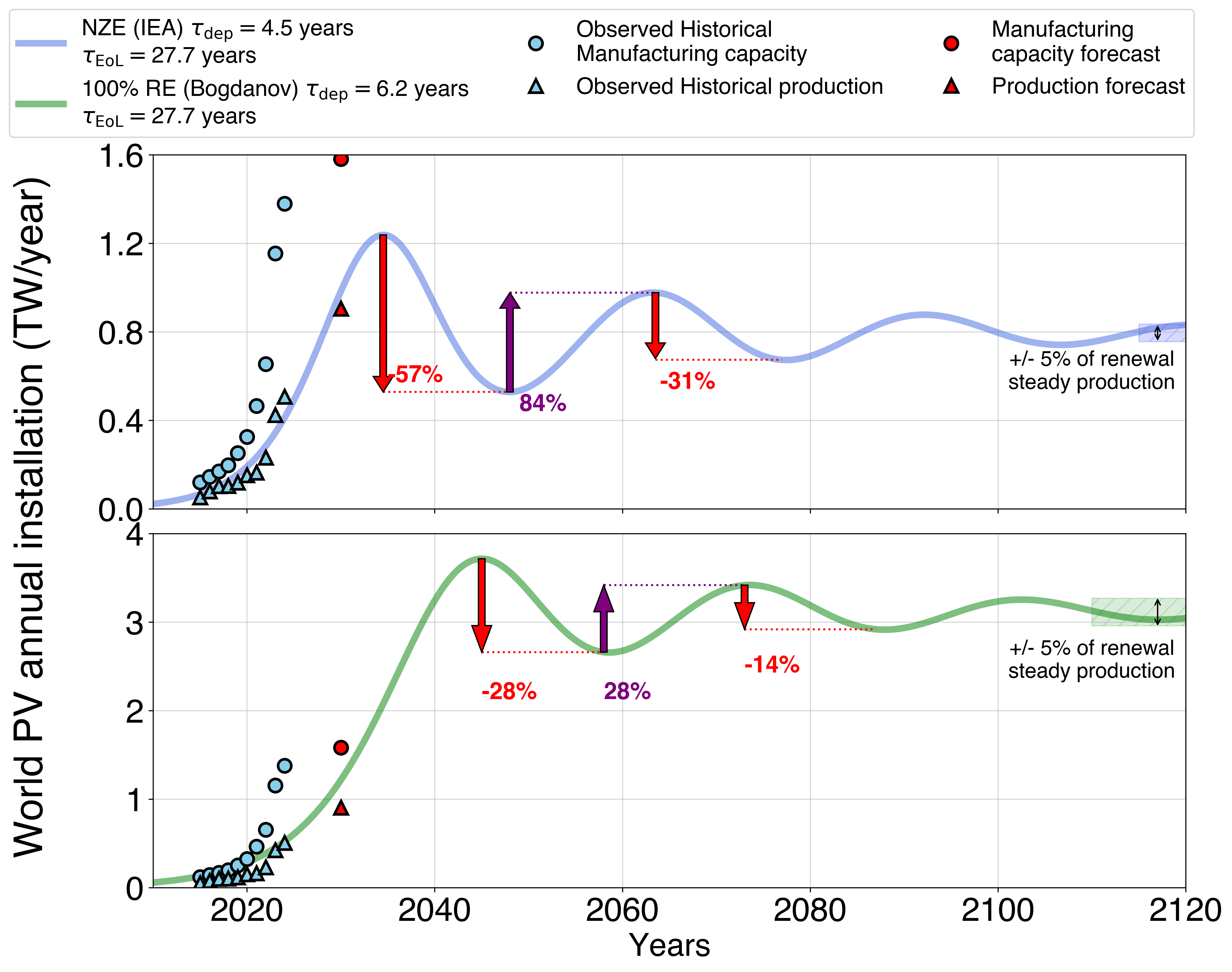}
    \caption{{\bf Solar panel production dynamics under different deployment scenarios.} This figure illustrates the production dynamics of solar panels for two deployment scenarios: NZE (blue) and 100RE (green) - see Table \ref{tab:PVscenarios} for the correspondence) - each considered under IRENA (2016) EoL distribution model (see Table \ref{tab:PVEoLDistrib}). The solid lines corresponds to the annual panel production derived from the model. Blue triangles and circle corresponds to historical data for panel production and panel manufacturing capacity [\cite{iea2023a}]. Red triangles and circles corresponds to IEA forecasts regarding panel production and panel manufacturing capacity [\cite{iea2023a}]. The colored arrows emphasize the intensity of the oscillations in both scenario.The numerical values represent the variation between both ends of the arrow, expressed as a fraction of the production at the arrow's base. The time scale has been extended to 2120 to observe the stabilization of production.}
    \label{fig:PVDeployment}
\end{figure*}

\subsection{The case of PV}

Concerning the deployment of PV, we can estimate the parameters $\tau_{\text{dep}}$ and $\tau_{\text{EoL}}$ using data from section \ref{modeldata}. The characteristic deployment time $\tau_{\text{dep}}$ is mathematically defined via the formal logistic expression. However, this definition is not particularly intuitive. In practical terms, it measures the time taken to transition from $x\%$ of the target installed capacity to $(100-x)\%$. For all practical purposes, it can be considered a measure of the time at which the target capacity is reached.

Most scenarios achieve their target capacity by 2050, resulting in characteristic times that fall within a relatively narrow range of $4$ to $6$ years. In contrast, the two scenarios proposed by Shell reach their target capacity well after 2050, leading to significantly higher characteristic times. Notably, the final capacity level does not influence the dynamics: whether the target capacity is 10 TW or 100 TW, as long as it is achieved within the same timeframe and maintained thereafter, the characteristic time remains unchanged.

Regarding the average lifespan of panels, $\tau_{\text{EoL}}$, there appears to be a consensus in the literature around a value of 30 years. This aligns with the warranties provided by panel manufacturers, which typically range between 25 and 30 years [\cite{vdma}]. However, the distribution proposed by Tan, which halves $\tau_{\text{EoL}}$, would significantly impact the critical ratio $\tau_{\text{dep}} / \tau_{\text{EoL}}$, which is discussed below.

For PV systems, the ratio $\tau_{\text{dep}} / \tau_{\text{EoL}}$ is approximately $0.15$-$0.2$, which is below the threshold of $0.3$. This results in oscillatory behavior or a so-called \textbf{fast deployment} regime. The production scenarios corresponding to the NZE, 100RE deployment scenario, and an average lifespan of 28 years [\cite{Irena2016}] can be seen in Figure \ref{fig:PVDeployment}.

To ensure readability and clarity, we have chosen to present only the NZE and 100RE deployment scenarios. The disparity in target capacities (22 TW vs. 86 TW) substantially influences the annual replacement production level to which the system eventually converges (0.8 TW/year vs. 3 TW/year). However, this difference does not alter the overall dynamics, as both scenarios display pronounced oscillations in their production curves.

Conversely, the small variation in deployment timelines significantly affects the characteristics of the oscillations. In the 100RE scenario, where the target capacity is reached closer to 2060, the oscillations are notably less pronounced.

\begin{figure*}[!h]
    \centering
    \includegraphics[width=0.9\linewidth]{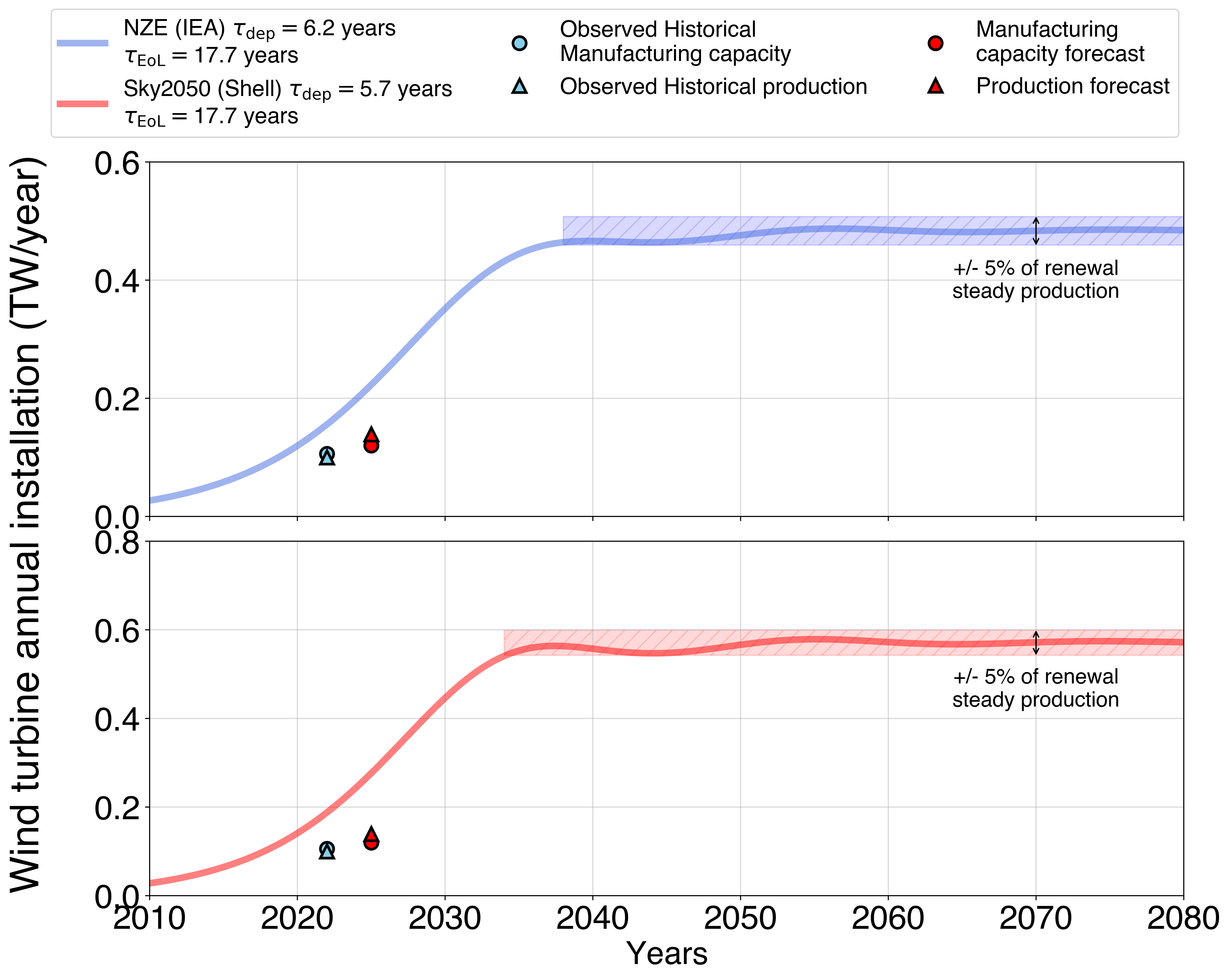}
    \caption{{\bf Wind turbine production dynamics under different deployment scenarios.} This figure illustrates the production dynamics of wind turbines for two deployment scenarios: NZE (blue) and Sky2050 (red) - see Table \ref{tab:PVscenarios} for the correspondence) - each considered under Cao (2019) EoL distribution model (see Table \ref{tab:WindEoLDistrib}). The solid lines corresponds to the annual turbine production derived from the model. Blue triangles and circle corresponds to historical data for turbine production and turbine manufacturing capacity [\cite{iea2023b}]. Red triangles and circles corresponds to IEA forecasts regarding turbine production and turbine manufacturing capacity [\cite{iea2023b}]. The model appears to overestimate wind energy production in 2022, which can be attributed to at least two factors. The first, detailed in \cite{LeBihan2025}, relates to the stochastic nature of the modeling approach. The model generates an expected production value rather than actual production, making comparisons with real-world data more relevant over longer time scales. Additionally, another possible explanation is the underestimation of wind turbine lifetimes, which would lead to an overestimation of replacement rates and, consequently, an overestimation of total annual production.} 
    
    \label{fig:WindDeployment}
\end{figure*}

\subsection{The case of Wind Power}

Following the same approach as for PV systems, the data presented in section \ref{modeldata} allow us to estimate the key parameters governing the global deployment of wind energy. Regarding the characteristic deployment time - $\tau_{\text{dep}}$ - most scenarios suggest a value in the range of $6$ to $7$ years. This is slightly higher than that observed for PV, indicating a somewhat slower deployment rate. This difference can be attributed to the fact that the installed wind capacity is already significant relative to the target value, which, as explained in the previous section, influences the logistic definition of the characteristic deployment time. Concerning the average lifespan of wind turbines - $\tau_{\text{EoL}}$ - there is a clear consensus on a range of $15$ to $20$ years. Consequently, the critical ratio $\tau_{\text{dep}} / \tau_{\text{EoL}}$ for wind energy falls within the range of $0.3$ to $0.47$, exceeding the $0.3$ threshold. As a result, unlike PV, wind energy follows a monotonic, non-oscillatory production regime, referred to as the \textbf{slow deployment} regime. The production scenarios corresponding to the NZE and Sky2050 scenarios, assuming an average lifetime of $18$ years [\cite{cao2019}], are illustrated in Figure \ref{fig:WindDeployment}.

The dynamics of these two scenarios are highly similar. A slight variation is observed in the renewal value to which the system converges, as the target capacity in the Sky2050 scenario is marginally higher than in the NZE scenario. The Sky2050 scenario with the shortest deployment time is close to the $0.3$ threshold for the $\tau_{\text{dep}} / \tau_{\text{EoL}}$ ratio ($5.7 / 17.7 = 0.32$), resulting in a minor decline in production post-peak before a subsequent increase. Only the 100RE scenario proposed a sufficiently short characteristic deployment time for wind energy to induce oscillations. However, given the discrepancies between the projections of this scenario and the actual installations observed between 2019 and 2024, its analysis for wind energy has been discontinued.

PV and wind energy thus provide examples of qualitatively distinct deployment regimes. The monotonically increasing wind energy production profile, which stabilizes at a renewal value, does not appear to introduce significant industrial constraints. In contrast, the oscillatory production patterns observed in PV panel deployment raise concerns about system stability and supply chain resilience. Since rapid deployment scenarios are the ones that pose the most pressing questions regarding industrial sustainability and long-term viability, an important part of this study focuses on analyzing these oscillations in detail. The next section presents a detailed analysis of the key indicators that define this oscillatory regime and examines their relationship with the dynamic parameters of deployment and end-of-life.

\subsection{Characterizing rapid deployment dynamics}

The various scenarios foresee the installation of more than $70\%$ of the PV target capacity within the average panel lifespan. This requires a production overshoot, leading to oscillations over an extended transition period. Here, we analyze the effects of deployment speed and average panel lifespan on this behavior, specifically focusing on production overshoot, the amplitude of production oscillations, and the duration of the transition phase.

\subsubsection{Production overshoot}

Production overshoot is defined here as the ratio of the production peak to the renewal level production. This ratio represents the additional production required to rapidly achieve the target capacity.

The production overshoot increases proportionally to the ratio $\tau_{\text{dep}}/\tau_{\text{EoL}}$. This proportionality arises from two core principles:

\begin{itemize}
    \item The production peak is inversely related to the deployment duration. In general terms, installing $K$ TW over a period of $\tau_{\text{dep}}$ years requires an average annual installation rate of approximately $K/\tau_{\text{dep}}$.
    
    \item The replacement production rate is inversely proportional to the average panel lifespan. As panels require replacement every $\tau_{\text{EoL}}$ years, the annual replacement rate for a total installed capacity of $K$ TW is given by $K/\tau_{\text{EoL}}$.
    
\end{itemize}

These two production rates arise from distinct constraints: one is determined by the timeframe available to achieve the target capacity, while the other is dictated by the lifespan of the panels. Deployment-driven production can greatly surpass renewal-driven production when the deployment period is short compared to the timescale defined by the panel lifespan and their corresponding renewal needs.

This ratio highlights how these two productions compare with each other. For oscillating or 'fast deployment' regimes scenarios, this value exceeds 100\% : as the accelerated pace of deployment demands a deployment production grater than the renewal one.

With a fixed panel lifespan, this phenomenon can be interpreted as a trade-off: achieving the target capacity, which is sustained over the long term, more rapidly results in higher overproduction. This trade-off will be explored further in the discussion, as it reflects the tension between the need for rapid deployment to meet climate goals and the challenges posed by overproduction from an industrial perspective. 

\begin{figure}[!h]
    \centering
    \includegraphics[width=\linewidth]{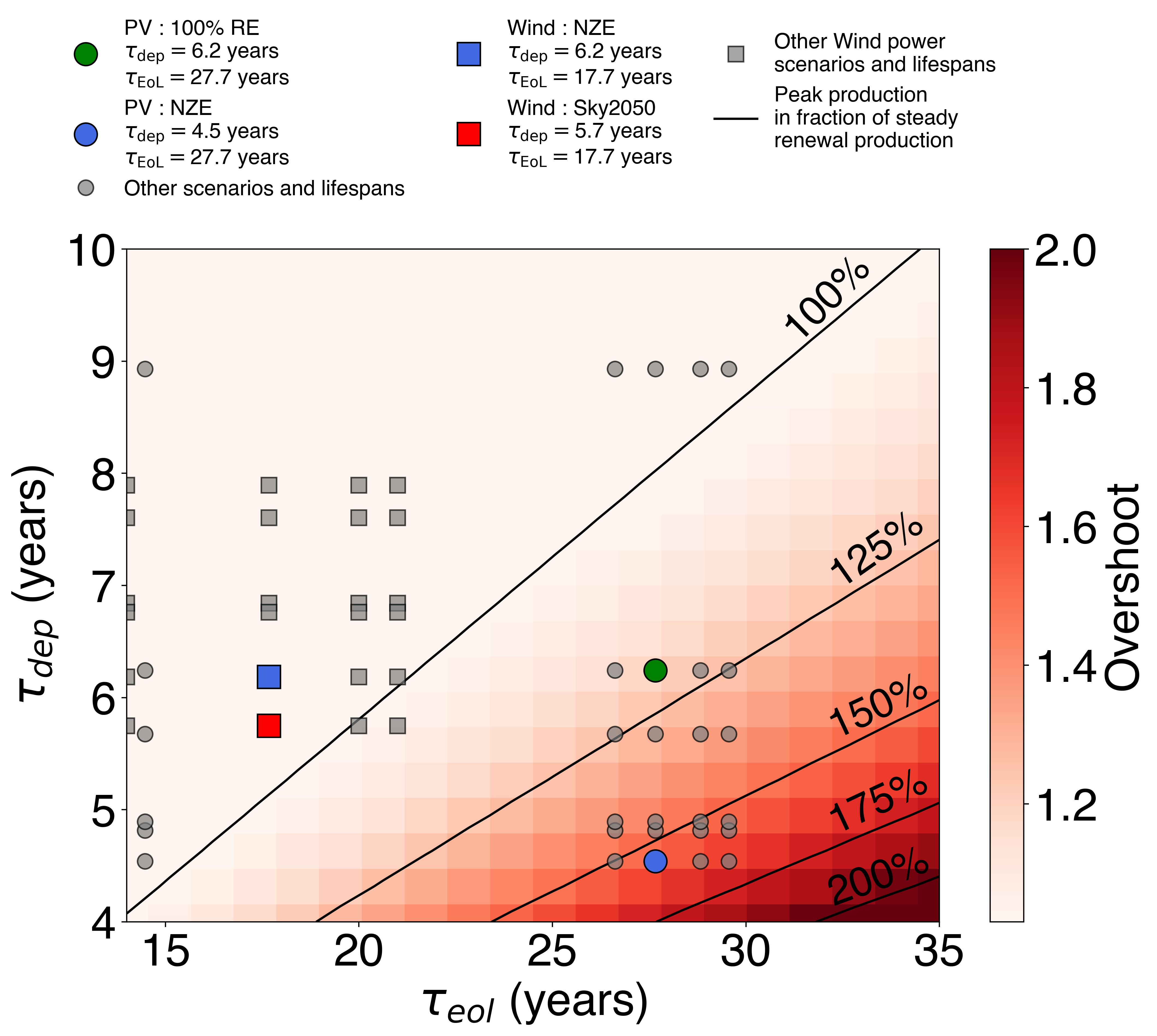}
    \caption{{\bf Relation between production overshoot, deployment rate and panel lifespan}. This figure shows production overshoot as a function of $\tau_{\text{dep}}$ values (linked to the target capacity attainment date) and $\tau_{\text{EoL}}$ values (panel average lifespan). The ($\tau_{\text{dep}}$, $\tau_{\text{EoL}}$) pairs are plotted for both solar and wind energy. It is observed that the data points corresponding to wind energy fall within the no-overproduction region, whereas those for solar energy can exhibit up to 150\% overshoot.}
    \label{fig:overprod}
\end{figure}

Figure \ref{fig:overprod} quantitatively illustrates the trade-offs between deployment speed, panel lifespan, and overproduction, highlighting the relationship between $\tau_{\text{dep}}$, $\tau_{\text{EoL}}$, and production overshoot.

The upper-left region of the figure \ref{fig:overprod} (white) represents the “slow deployment” regime, where the characteristic deployment time is sufficiently long relative to the average panel lifespan, allowing production to increase steadily without overshoot. Conversely, the lower-right region (various shades of red) beneath the “100\%” curve reflects the overshoot regime: the short timeframe required to reach the target capacity requires overproduction. The $100\%$ line corresponds to the $\tau_{\text{dep}}/\tau_{\text{EoL}}=0.3$ threshold, which separates oscillatory behavior from monotonic growth.

Overproduction is clearly proportional to the ratio $\tau_{\text{dep}}/\tau_{\text{EoL}}$: the farther you move diagonally towards the bottom right corner, the greater the overshoot. For the NZE scenario, this overshoot reaches more than 150\%.

\subsubsection{Oscillations in production}

The overproduction metric introduced in the previous section serves as a long-term indicator, quantifying the magnitude of the difference between deployment-driven production and the steady-state renewal level. However, this steady-state renewal level may only be reached several decades after deployment. Consequently, this long-term indicator does not fully capture the fluctuations that can arise during this transition period. Although the observed oscillations are partly a consequence of  initial overproduction, they are also significantly influenced by the variability in panel lifespans.

As discussed in Section \ref{sec:twodynamics}, production undergoes oscillations driven by replacement waves. These oscillations are illustrated in Figure \ref{fig:PVDeployment}. The first red arrow indicates the decline in production as capacity saturates and additional panels are no longer needed. The purple arrow marks the subsequent increase in production, triggered by a wave of panels reaching the end of their lifespan. These fluctuations are pronounced and occur over relatively short time intervals.

The presence and intensity of these oscillations, like overproduction, depend on the ratio $\tau_{\text{dep}} / \tau_{\text{EoL}}$. In addition, they are influenced by the concentration of the lifespan distribution, which is inversely proportional to the coefficient of variation $CV_{EoL}$. A more concentrated distribution (typically $CV_{EoL} < 0.3$) results in more synchronized end-of-life events, leading to a more pronounced replacement peak. In contrast, higher $CV_{EoL}$ values (typically $CV_{EoL} > 0.3$) reflect desynchronized end-of-life events, reducing the intensity of the replacement peak.

\begin{figure*}[!h]
    \centering
    \includegraphics[width=\linewidth]{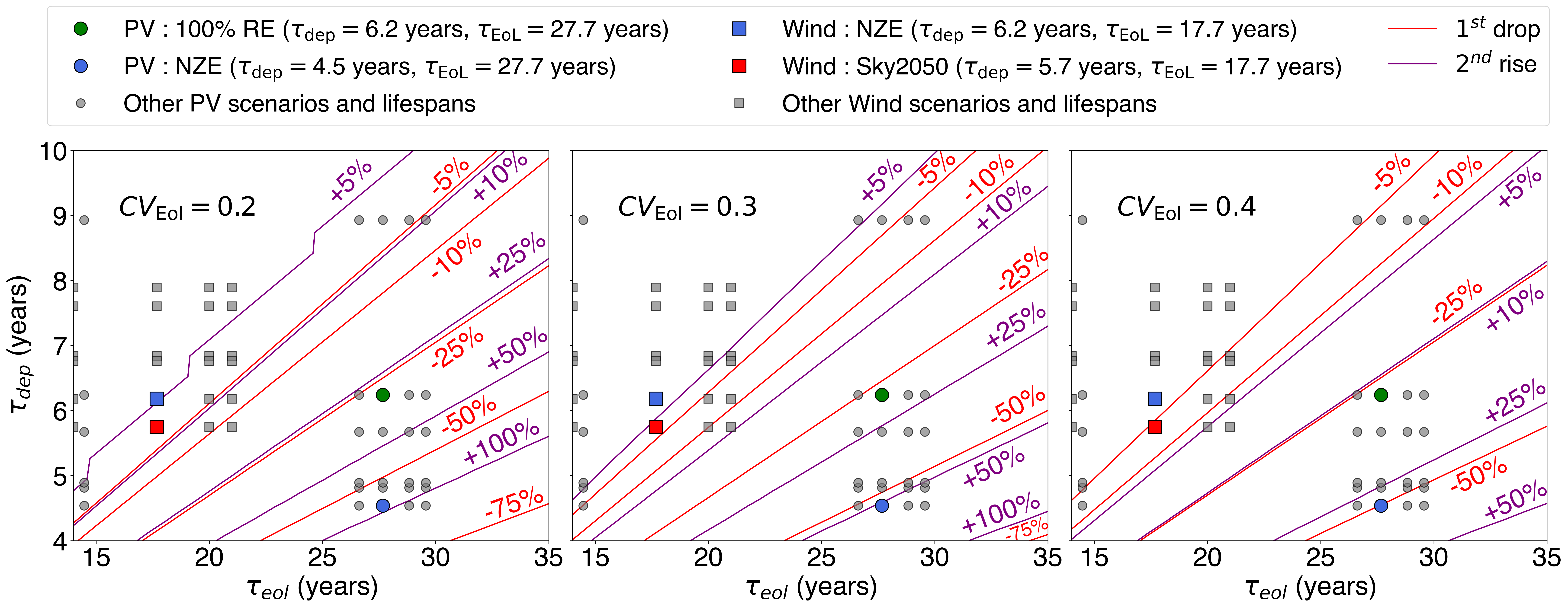}    
    \caption{{\bf Relation between oscillation intensity, deployment rate and equipment lifespan}. This figure illustrates production oscillations as a function of the deployment rate ($\tau_{\text{dep}}$ in years) and the equipment average lifespan ($\tau_{\text{EoL}}$ in years) for different dispersions of equipment lifetime distributions.  The different solar scenarios and average panel lifespans are displayed, along with the corresponding data for wind power. The graph can be interpreted as follows: the red lines represent the decrease in production between the peak and the first trough. The purple lines  correspond to the increase in production between the first trough and the second peak, referred to as the replacement peak. These variations are expressed as a percentage of production (see red and purple arrows in Figure~\ref{fig:PVDeployment}). The lines delineate regions where the variation is less than the value indicated by the line (above the line) versus greater (below the line). The blue and green dots represent the NZE and 100RE scenarios, respectively. For instance, with low dispersion (\(CV_{\text{eol}} = 0.2\)), the production increase for the NZE scenario is nearly 100\%, meaning the volume of panels produced must double to accommodate the replacement peak. In contrast, for higher dispersion (\(CV_{\text{eol}} = 0.4\)), the increase is only $25\%$. In general, dispersion has little effect on the first decrease, which depends primarily on overproduction, but significantly influences the second rise: the more panels reaching the end of their life simultaneously, the greater the replacement peak. The blue and red squares, representing the NZE and Sky2050 scenarios for wind energy, are located in the parameter space corresponding to minimal or no variations in production.}
    \label{fig:oscillations}
\end{figure*}

Figure \ref{fig:oscillations} illustrates that the replacement peak (the second increase in purple corresponding to the purple arrow in Figure \ref{fig:PVDeployment}) is lower for $CV_{EoL}=0.4$ (right panel) compared to $CV_{EoL}=0.3$ (center panel), which in turn is lower than $CV_{EoL}=0.2$ (left panel). Similar to the previous figure, this visualization enables the positioning of different (deployment scenario, lifespan distribution) pairs and provides a quantitative measure of the oscillation levels.

Similarly, the combination of (NZE, Irena) (associated with a $CV_{EoL}$ of $0.2$) results in significant fluctuations: an initial decline by 50\%, followed by an 80\% increase, all occurring within a time-frame of less than twenty years.

\subsubsection{Transition duration}

\begin{figure}[!h]
    \centering
    \includegraphics[width=\linewidth]{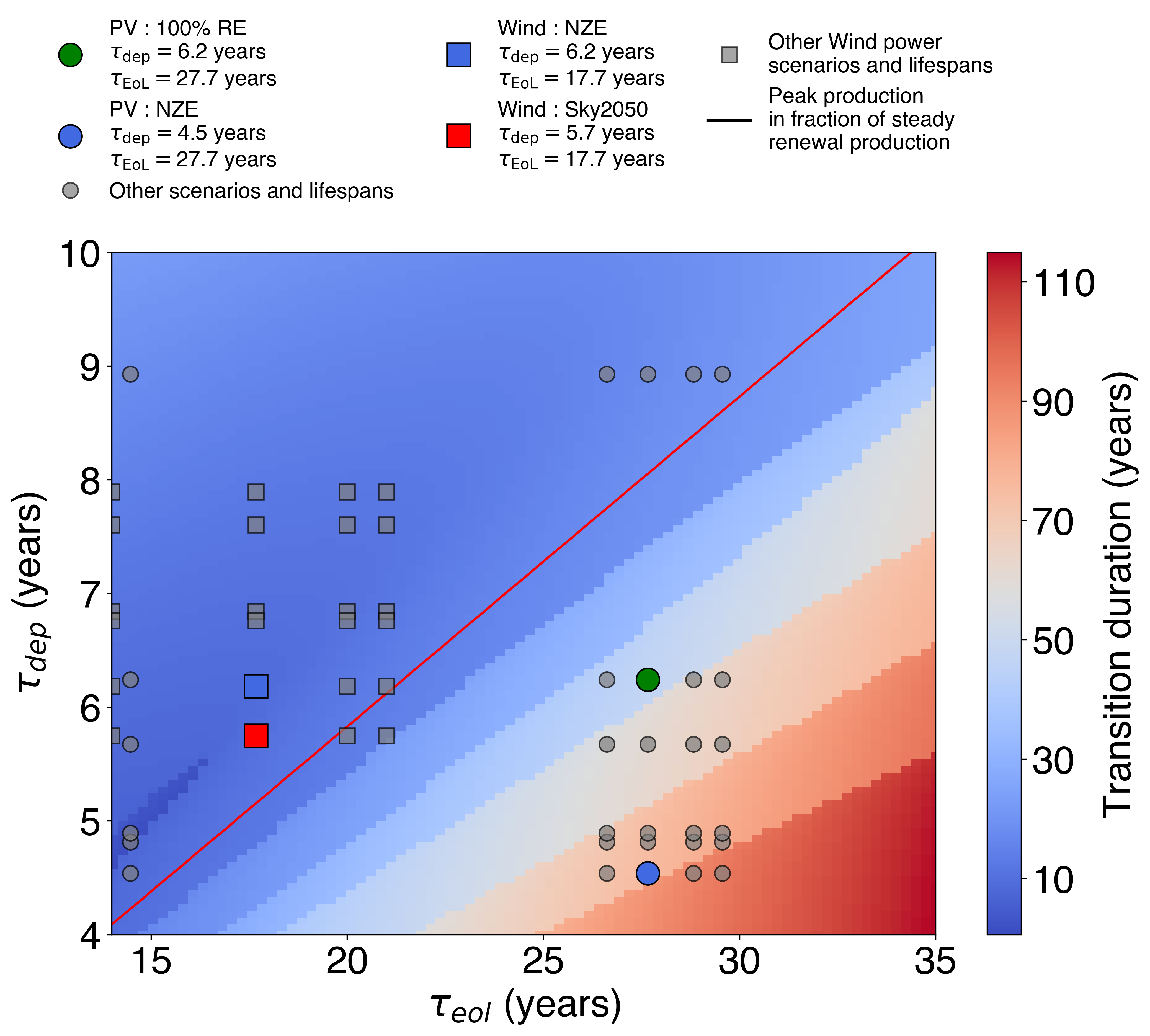}
    \caption{{\bf Relation between transition duration, deployment rate and equipment lifespan}. This figure shows transition time as a function of $\tau_{\text{dep}}$ values (linked to the target capacity attainment date) and $\tau_{\text{EoL}}$ values (equipment average lifespan). Time to reach the Renewal Steady Production for different scenarios and different lifetime distributions. Transition duration is defined here as the time laps between the peak production and the date at which the production oscillations are less than $5\%$ of the steady renewal value. This $5\%$ threshold is a parameter itself that affects the results. That is what creates artificial frontiers between color bands, corresponding to the parameter couple for which nth oscillation just exceed the $5\%$ threshold.}
    \label{fig:transitionduration}
\end{figure}

Finally, an essential aspect of the oscillatory transition period (highlighted in yellow in Figure \ref{fig:Intro}) is its duration and how it compares to the deployment time. This duration is inherently linked to the intensity of the oscillations. As with Figures \ref{fig:overprod} and \ref{fig:oscillations}, Figure \ref{fig:transitionduration} illustrates the transition times for various combinations of ($\tau_{\text{dep}}$, $\tau_{\text{EoL}}$). Specifically, it shows the time span between the production peak and the stabilization of production around its renewal value.

Unlike the production peak, the point of stabilization depends on how stabilization is defined. Here, stabilization is considered achieved when production deviations from the renewal value remain within 5\% of that value. This threshold is arbitrary; increasing it would naturally result in shorter transition durations.

In general, regardless of the threshold, the further below the red line (indicating the onset of oscillations), the longer the transition period. To grasp the effect of this arbitrary threshold, one can consider Figure \ref{fig:PVDeployment} and imagine a wider rectangle. For the 5\% threshold, production in the NZE scenario does not stabilize until nearly 70 years after the peak, meaning that stabilization is not achieved until the early 22nd century.

\section{Discussion}

The production dynamics associated with wind energy, characterized by slow deployment, and solar photovoltaics, characterized by fast deployment, are qualitatively distinct. The slow deployment scenario does not impose industrial constraints and is fully compatible with the gradual scaling of an industry. In contrast, fast deployment scenarios, associated with production overshoot and industrial oscillations, are challenging from a policy perspective.

\subsection{Strategic implications for a global photovoltaic system}

Fast deployment scenarios imply major alternating changes in the subsystems with fast structuring and de-structuring of the PV value chain.  
Public policies and industry strategies must take this into account and design adapted policy frameworks that are able to accompany industries through the deployment to the renewal phases and provide short-term and long-term policy framework and planning guidelines.

The following sections provide an initial approach to characterizing the risks associated with a rapid deployment regime that PV technology may experience. It is important to keep in mind the generality of these considerations and their applicability to other technologies, particularly those involved in the energy transition, that may face similar dynamics. For instance, this could also apply to wind energy if turbine lifetimes were extended to 30 years.

\subsubsection{How to define the PV technological system}

This study has focused on the deployment and renewal dynamics of the solar panel fleet, which is considered a tangible technological object. However, one must bear in mind that behind this technological item lies a complex set of societal dynamics composed of material and non-material subsystems including: mineral supply, factories, labor force, regulations, markets, knowledge, etc. In the literature on social sciences, this inherent complexity is referred to as 'sociotechnical system' [Mumford (1934), Simondon (1958), Gras (1997), Schatzki (2019), Raineau (2022)]. This concept suggests that a solar panel – as all technological items – is embedded into a broad range of complex social, economic and technical interactions, which together form: the PV technological system. 
The PV technology is currently considered a potential solution in the energy transition because these specific sociotechnical configurations hold together. But this brings us to an essential strategic issue: aiming for a renewal regime for photovoltaic capacity raises the question of the renewability of all the subsystems on which the solar panel industry relies. In other words, renewing solar PV capacity questions the renewal of all the sociotechnical configurations sustaining the PV system.

\subsubsection{Strategic issues for PV renewal}

Fast-deployment scenarios, characterized by production overshoots and industrial oscillations, present significant challenges from a policy perspective. These scenarios would involve structural changes throughout the global photovoltaic value chain, leading to rapid restructuring and disruption dynamics. 

From an economic point of view, production overshoot represents a major risk in terms of stranded assets. This risk has been studied in the context of the energy transition [\cite{chick2007,caldecott2017,vondulong2023,chaudary2024}] but has not yet been applied to the technological system of photovoltaics from a long-term perspective. As our model indicates, oscillating scenarios are highly probable for PV, making the risk of stranded assets significant. This risk mainly arises from excess production capacity during the overshoot phase, which exceeds the target capacity. If left unaddressed, these dynamics could lead to structural lock-ins within the PV industries and create long-term path dependencies on growth [\cite{unruh2000}]. These industries could be locked into unsustainable practices, with the most likely planned obsolescence example, to avoid the asset losses induced by overshoot. 

From a labor perspective, the oscillations triggered by production overshoots may have far-reaching implications. According to a recent report from the IEA, the PV industry could create 1300 manufacturing jobs for each gigawatt of production capacity until 2030 [\cite{iea2022}]. However, our projections of the IEA NZE and Bogdanov scenarios show a drop of around 1 TW/year of PV production capacity between the overshoot and the beginning of the first renewal wave. This drop could result in the loss of almost one million jobs, underscoring the need for careful policy planning and mitigation strategies. 

In addition, the relatively long lifespan of solar PV systems raises critical issues around skill transfer. Short-lifespan technologies are typically built and renewed by the same generation of workers. However, with longer-lifespan technologies such as PV, initial deployment and subsequent renewal are likely to involve different generations of workers, necessitating the transfer of specialized skills. This challenge is similar to what has been observed in nuclear power plant operations [\cite{LeBihan2025}]. Given that solar panel deployment tends to follow oscillating patterns, it would be prudent to explore the interoperability of solar PV labor with adjacent sectors, such as metallurgy, manufacturing, construction, etc., to ensure adaptability of the  PV workforce. 

From a geopolitical perspective, oscillating scenarios introduce additional challenges, particularly with regard to international relations and supply chain security. In these scenarios, the discontinuity of PV demand raises questions about the long-term availability of the different PV components for renewal. The high market concentration of manufacturing stages in a single country (China) represents additional challenges in this regard. In fact, between 2010 and 2022, China further strengthened its dominant position in all stages of photovoltaic manufacturing, including polysilicon production, wafer, cell, and module manufacturing. The Chinese industry currently dominates over 80\% of global capacity in all segments of the PV value chain. Furthermore, the intricate interconnection between photovoltaic production and raw material supply creates substantial long-term geopolitical vulnerabilities in resource security [\cite{iea2024c,fizaine2015}].

To effectively address these complex challenges, policymakers must establish regulatory frameworks that guide industries through both the deployment and renewal phases of the photovoltaic (PV) technological system. This requires an integrated approach, combining short-term strategies to address immediate constraints with long-term policies that offer clear planning directives, ensuring a seamless transition from deployment to renewal dynamics.

\subsection{About the time scales that should be considered}

By extending PV deployment scenarios beyond the conventional 2050 horizon, we have highlighted the critical importance of addressing panel renewal dynamics. After 2050, the production of panels to replace those reaching their EoL becomes the primary, if not the sole, driver of PV manufacturing. Before delving into a methodological framework discussion, two general yet essential insights can be emphasized. 

First, once a significant number of panels reach their EoL —~where "significant" refers to a scale comparable to the annual panel production~— any increase in active capacity (i.e. the number of operational panels) will necessarily require an increase in production capacity. This is because a fraction of manufactured panels will be allocated solely to replacements rather than expand the installed capacity. 

Second, the timing of this transition is determined by the deployment trajectory and the effective lifetime of PV panels. The exponential growth observed in PV capacity deployment suggests that if the installed capacity saturates before the production capacity, the industry will inevitably face the deployment-renewal transition described here. This situation is plausible given that the constraints on installed capacity, driven by factors such as climate targets and energy policy, differ fundamentally from those on production capacity. 

A historical precedent for this dynamic can be found in the nuclear energy sector, where deployment followed political imperatives [\cite{LeBihan2025}]. Renewable energy technologies, particularly PV, exhibit numerous characteristics analogous to this pattern, suggesting that similar challenges may arise during their long-term deployment and replacement phases.

\subsection{What Does This Model Teach Us?}

The model presented in this study provides a first-order approximation of the dynamics of deploying and renewing a large-scale solar and wind farm. It is based on several assumptions that naturally influence the analyses derived from it. In this final part of the discussion, we revisit the insights offered by a simplified model like this one, as well as its limitations, which could be refined to enhance its applicability to real-world scenarios.

\subsubsection{Key Insight}

The primary insight from this model is the emergence of oscillations that can arise endogenously during the deployment and long-term renewal of a technology. This phenomenon appears to be characteristic of the global deployment of PV systems, whereas it does not apply to the global deployment of wind energy. These oscillations become more pronounced when the ratio between the deployment time and the lifespan of the panels is low. The main takeaway is that the transition from a deployment-driven to a renewal-driven dynamic may induce a significant transitional period.

\subsubsection{Relevance to the photovoltaic technology system}

This finding is particularly relevant for PV systems for two main reasons. First, most data from the literature (scenarios and lifespans) suggest that PV deployment is likely to experience an oscillatory regime. A natural first step would be to increase research efforts focused on understanding real-world panel lifespan distributions and deployment speeds, rather than solely on final capacity targets outlined in various scenarios.

More broadly, the oscillatory regime could pose challenges for all RE technologies for two structural reasons. First, the drive to accelerate the energy transition tends to shorten the characteristic deployment time of technologies, effectively steepening the deployment curve of active capacity. Second, in line with resource conservation, sustainability, and the minimization of emissions associated with panel production, the effective lifespan of panels and other RE infrastructures is expected to increase. For PV panels, this trend is already evident.

These two effects structurally contribute to the emergence of oscillations. The model highlights a tension: one potential solution to avoid oscillations would be to slow deployment and reduce panel lifespans. However, while lifespan dispersion might mitigate oscillation amplitudes through complex industrial strategies, this does not provide a straightforward solution. The takeaway is not the desirability of a monotonic regime due to reduced industrial risk, but rather the recognition of the risks posed by the oscillatory regime, which should concern all RE technologies. Given that this oscillatory regime is deeply intertwined with the goals of a rapid and resource-efficient energy transition, this paper raises questions about the real-world implications of such phenomena and strategies for managing them, rather than merely how to adjust parameters to avoid them.

\subsubsection{Revisiting Model Assumptions}

A crucial assumption in this model is the exogeneity of active capacity. This constraint is predefined based on a given scenario. While this allows for interrogation of the scenario itself, does it enable the model to provide a realistic production scenario? This largely depends on the real-world dynamics of constraints. In the model, production adjusts to meet the deployment of the target capacity. However, for some technologies, the relationship could be reversed—for instance, when demand greatly exceeds production capacity. In such cases, production saturates naturally (no oscillations), and the growth of active capacity is capped. Similarly, panel lifespans could adapt to production constraints: consumers may replace equipment with new generations irrespective of technical end-of-life. This adaptive behavior is partially addressed in \cite{Tan2022}, showing that it results in a monotonic deployment regime (gray points in the left side of Figure \ref{fig:overprod} to \ref{fig:transitionduration} corresponding to the shortest average lifespan of ).

However, PV systems have specific features —~such as political deployment goals, efforts to extend lifespans, and the difficulty for individuals to replace installations~— that challenge the nature of the constraint. More interdisciplinary studies are necessary to capture the complexity of these constraint relationships. Such efforts will be crucial for the design of large-scale public policies and industrial strategies.

\section*{Conclusion}

The streamlined model developed in this study sheds light on key challenges related to the deployment and maintenance of a large-scale global solar and wind park. It emphasizes the significant demand for both panel and wind turbines production required for renewal, a factor that cannot be overlooked. In particular, even as we approach 2050, current annual production levels fall short of meeting the renewal needs necessary to sustain the projected global capacity.

A key insight from this study is the trade-off between accelerating capacity deployment and the resulting overproduction. In scenarios like the IEA’s Net Zero Emissions, solar overproduction could reach up to 150\%, which means that one-third of the production capacity established during the peak around 2035 may become unsustainable in the long term. This presents significant challenges for industrial planning and investment strategies.

In addition to long-term constraints, the model highlights medium-term challenges arising from production fluctuations after the deployment peak. Rapid deployment can amplify these fluctuations, resulting in sharp declines in production and pronounced replacement surges. These disruptions create significant weaknesses in the system that specifically affect investment stability, retention of workforce skills, and resilience of supply chains.These challenges appear to be primarily relevant to the photovoltaic industry, as wind energy deployment results in a monotonic production pattern.

Finally, the model suggests potential mitigation strategies, particularly by utilizing the variability in the lifespans of panels to stabilize production dynamics. These findings lay the groundwork for more detailed investigations and emphasize the importance of proactive measures to address the industrial and economic challenges associated with the deployment and renewal of large-scale RE capacity.

The model is useful for analyzing the material demands required to achieve the scenarios and the potential fluctuations associated with them. This will be the focus of our next related study.

 \section*{CRediT authorship contribution statement}
 
\textbf{Joseph Le Bihan:} Investigation, Methodology, Formal analysis, Conceptualization, Data curation, Visualization, Programming, Implementation of the computer code and supporting algorithms, Writing - Original Draft.

\textbf{Thomas Lapi:} Investigation, Methodology, Writing - Review \& Editing.

\textbf{José Halloy:} Conceptualization, Methodology, Investigation, Formal analysis, Writing - Review \& Editing, Supervision, Project administration, Funding acquisition.

\section*{Declaration of competing interest}

All authors declare that they have no conflicts of interest.

\section*{Data access}

All code and data used in this study are available on Zenodo,
DOI 10.5281/zenodo.14803331

\section*{Acknowledgements}
  This work has benefited from a government grant managed by the Agence Nationale de la Recherche under the France 2030 program, reference ANR-22-PERE-0003.





\end{document}